\documentclass{article}
\usepackage{spconf,amsmath,graphicx,hyperref}

\usepackage{commands}

\usepackage{tikz}
\usetikzlibrary{positioning, calc}

\title{Strong Basin of Attraction for Unmixing Kernels With the\\Variable Projection Method}

\name{Santos Michelena$^{\star \dagger}$ \qquad Maxime Ferreira Da Costa$^{\star}$ \qquad José Picheral$^{\star}$}
  
\address{$^{\star}$ Laboratory of Signals and Systems, CentraleSupélec, Université Paris–Saclay, CNRS, France\\
      $^{\dagger}$ Iumtek, Orsay, France} 

\begin{document}

\maketitle

\begin{abstract}
    The problem of recovering a mixture of spike signals convolved with distinct point spread functions (PSFs) lying on a parametric manifold, under the assumption that the spike locations are known, is studied. The PSF unmixing problem is formulated as a \emph{projected non-linear least squares estimator}.
    A lower bound on the radius of the region of strong convexity is established in the presence of noise as a function of the manifold coherence and Lipschitz properties, guaranteeing convergence and stability of the optimization program. Numerical experiments highlight the speed of decay of the PSF class in the problem's conditioning and confirm theoretical findings. Finally, the proposed estimator is deployed on real-world spectroscopic data from \emph{laser-induced breakdown spectroscopy (LIBS)}, removing the need for manual calibration and validating the method’s practical relevance.
\end{abstract}

\begin{keywords}
non-convex optimization; PSF unmixing; manifold constraint; variable projection method; laser-induced breakdown spectroscopy
\end{keywords}

\section{Introduction}

Point spread function (PSF) unmixing is the problem of recovering a mixture of spike signals after convolution with group-dependent PSFs. Such parametric models arise in several areas of applied science and engineering, including super-resolution imaging~\cite{huang2008three,huang2018single}, neural spike sorting~\cite{knudson2014inferring,li2014sparse}, multi-path channel identification in wireless communications~\cite{applebaum2012asynchronous,chi2013compressive}, and calibration-free spectroscopy~\cite{review_cf_libs}.  

While blind deconvolution for a single modality has been extensively studied~\cite{li2016identifiability}, much less is known about sparse mixtures with multiple, unknown PSFs. Convex approaches~\cite{romberg2010sparse,chi_guaranteed_2016,Chretien2020} provide stable recovery guarantees but constrain the PSFs to a predetermined low-dimensional subspace and are computationally costly. Non-convex methods scale better~\cite{traonmilin2024strong,costa_local_2023,gabet2025global}, yet they typically rely on restrictive assumptions, often tied to Fourier-domain models. These limitations make existing techniques ill-suited for applications such as laser-induced breakdown spectroscopy (LIBS), where PSF shapes reflect nonlinear plasma dynamics and do not conform to simple low-rank structures~\cite{mars-spectrum-fitting}.  

In~\cite{michelena}, PSF unmixing with known spike support is tackled via nonlinear least squares. Coherence and interference functions are introduced to control the loss geometry and estimate the convergence regions of first-order optimization methods. Our main contribution builds upon this framework to introduce a novel projected nonlinear least squares estimator based on the \emph{variable projection method} (VarPro)~\cite{variableProjection,varpro-meta-review,projected-hessian}.
The separable structure of the model is exploited through the use of the Moore–Penrose pseudo-inverse, thereby obtaining a reduced and tractable problem that is independent of the amplitude of the sources.
We establish in Theorem~\ref{thm:radius-projected} an explicit lower bound on the radius of convergence around the ground truth in the presence of noise.

The rest of the paper is organized as follows. In Section~\ref{sec:problem}, the PSF unmixing problem is presented, and the separable structure of the parameter is used to formulate a \emph{projected non-linear least square estimator}. Section~\ref{sec:main-result} presents our main result in Theorem~\ref{thm:radius-projected}, after introducing the notions of \emph{total coherence} and \emph{Lipschitz continuity}, which are needed to control the optimisation landscape. A sketch proof is proposed in Section~\ref{sec:proof}. Finally, numerical experiments are conducted in Section~\ref{sec:experiments},  confirming our theoretical findings. The practical relevance of the approach in LIBS spectroscopy is demonstrated, where it jointly estimates line widths and amplitudes, thereby removing the need for manual calibration and automating a key step in quantitative analysis.

\section{Problem Formulation}\label{sec:problem}

\paragraph*{Signal and measurement model.} We consider a parametric class of kernels, or PSFs, $g(\cdot,\cdot)$, where $t \mapsto g(\theta,t)$ is the PSF with parameter $\theta \in \Theta$. We assume the signal of interest $x(t)$ is a linear mixture of translated PSFs within a class. That is
\begin{equation}\label{eq:signal-model}
    x(t) = \sum_{i=1}^p \sum_{\ell=1}^{M_i} \eta^\star_{i,\ell}\, g(\theta_i^\star, t - t_{i,\ell}).
\end{equation}
The mixture number $p$, the spike counts $\{M_i\}$, and the source locations $\T \coloneq \{t_{i,\ell}\}$ are supposed to be known. The \emph{unknowns} of the problem are: the shape parameters of the mixture elements $\{\theta_i^\star\}$, and the sources amplitudes $\{\eta_{i,\ell}^\star\}$. 

The continuous signal $x(t)$ is sampled on $N$ uniformly spaced instants $\{u_s\}_{s=1}^N$ within the sampling interval $I=[-T/2,T/2]$. For any $\bm{\theta} \in \Theta^p$, we define the dictionary matrix
\begin{equation}
    \bG(\btheta) \coloneq [\bg_{1,1}, \cdots, \bg_{1, M_1}, \cdots, \bg_{1,p}, \cdots, \bg_{p,M_p}] \in \mathbb{R}^{N \times M},
\end{equation}
where $\bg_{i, \ell} \coloneq \big[g(\theta_i,u_s-t_{i,\ell})\big]_{s=1}^N \in \R^N$ and $M = \sum_{i=1}^p M_i$ is the model order. Furthermore, the amplitude vector is

\begin{equation}
    \etab^\star \coloneq [\eta_{1,1}^\star,\dots,\eta_{1,M_1}^\star,\dots,\eta_{p,1}^\star,\dots,\eta_{p,M_p}^\star]^\top \in \R^M.
\end{equation}
Finally, given additive noise noise $\bm{w}$, the observation $\bm{x}$ writes
\begin{equation}\label{eq:general-model}
    \bx = \bG(\btheta^\star)\etab^\star + \bm{w},
\end{equation}
where $\bx^\star = \bG(\btheta^\star)\etab^\star$ is the ground truth.  

For our analysis, we further assume that $N > M$ and that $\bG(\btheta)$ has full column rank for any $\bm{\theta} \in \Theta^p$. Therefore, its Moore–Penrose pseudo-inverse exists and is denoted $\bG(\btheta)^+ = \big(\bG(\btheta)^\top\bG(\btheta)\big)^{-1}\bG(\btheta)^\top \in \R^{M\times N}$. Furthermore, we assume $g(\cdot,t) \in \mathcal{C}^2(\Theta)$ for all $t \in I$, $g(\theta, \cdot)$ is even, and  $\partial_a g(\theta,\cdot) \in L^1(\R)\cap L^2(\R)$ for any $\theta \in \Theta$ and $a \in \{0,1,2\}$.

\paragraph*{The variable projection method.} Given an observation $\bm{x}$ following~\eqref{eq:general-model} the \emph{projected least squares estimator}~\cite{variableProjection} estimates parameter $\bm{\btheta}^\star$ with a minimizer $\hat{\bm{\theta}}$ of the loss
\begin{equation}\label{eq:projected-lstq}
    \mathcal{L}(\bm{\theta}) = \tfrac{1}{2N}\norm{\P_{\bG(\btheta)}^\perp\bx}_2^2,
\end{equation}
where $\P_{\bG(\btheta)}^\perp \coloneq \Id_N - \bG(\btheta)\bG(\btheta)^+$. 
The source amplitudes are subsequently estimated with $\hat{\bm{\eta}} = \bm{G}(\bm{\hat{\theta}})^{+} \bm{x}$.

Given the non-convex nature of the loss in~\eqref{eq:projected-lstq}, the \emph{local} convergence properties of the problem are studied. In particular, it is important to characterize the \emph{strong basin of attraction} for the loss~\eqref{eq:projected-lstq} in which the ground truth $\btheta^\star$ lies. That is, the neighborhood $\mathcal{N}$ of $\bm{\theta}^\star$ such that there exist a constant $0 < \xi$ such that $\L$ is $\xi$-strongly convex on $\mathcal{N}$. That is, for all $\btheta \in \mathcal{N}$ and all $\bu \in \R^p$,
\(
    \xi \norm{\bu}_2^2 \leq \bu^\top \nabla^2\L(\btheta)\bu.
\)
First-order optimization methods are guaranteed to converge towards a unique minimizer when initialized within the strong basin of attraction of $\bm{\theta}^\star$.
As our main contribution, we derive a lower bound for the size of the basin of attraction in terms of problem parameters.

\section{Convergence of the VarPro Estimator}\label{sec:main_result}

\subsection{Preliminaries}
\paragraph*{Minimal separation} First, we introduce the \emph{minimal separation} of the support $\mathcal{T}$, which entirely governs the conditioning of the dictionary matrix $\bm{G}(\btheta)$. It is defined as 
\begin{equation}
    \Delta \coloneqq \min_{t,t' \in \T, \; t \neq t'} |t - t'|.
\end{equation}

\paragraph*{Coherence and total coherence.} 
Next, we introduce the notions of \emph{coherence} and \emph{total coherence}, which extend classical notions frequently used in compressive sensing and sparse recovery, to parametric dictionary settings.

Herein, coherence measures the maximal correlation between parameter-dependent dictionary elements and their derivatives when separated by at least~$\Delta$. For $a \in \{0,1,2\}$, set
\begin{equation}
    \partial_a \bg(\theta, t) := \big[ \partial_a g(\theta, u_s - t) \big]_{s=1}^N \in \R^N.
\end{equation}
Then the \emph{coherence} between $\partial_a g(\theta_i,\cdot)$ and $\partial_a g(\theta_j,\cdot)$ at separation $\Delta$ is defined as
\begin{equation}\label{eq:def_coherence}
    \mu_{a}(\theta_i, \theta_j, \Delta) \coloneqq \sup_{|\delta|\geq \Delta} \left| \partial_a \bg(\theta_i, 0)^{\top} \partial_a \bg(\theta_j, \delta)  \right|.
\end{equation}
The \emph{total coherence} aggregates such correlations to quantify the similarity between parametric dictionary blocks:
\begin{align}
\label{eq:def_coherence_function}
\mathcal{C}_{a}(\theta_i, \theta_j, \Delta) \coloneqq \sum\limits_{m \in \Z\setminus\{0\}} \mu_{a}(\theta_i, \theta_j, |m|\Delta)
\end{align}

\vspace{-10pt}
\paragraph*{Lipschitz continuity.} To control the behavior of the Hessian in a neighborhood of $\btheta^\star$, we introduce Lipschitz constants for the coherence and the total coherence. Specifically, define $C_\mu \coloneqq \max_{a \in \{0,1,2\}} C_\mu^{a}$ and $C_\Delta \coloneqq \max_{a \in \{0,1,2\}} C_\Delta^{a}$, where $C_\mu^{a}$ and $C_\Delta^{a}$ are the Lipschitz constants of the maps 
$\theta \mapsto \mu_{a}(\theta,\theta',0)$ and 
$\theta \mapsto \C_{a}(\theta,\theta',\Delta)$, respectively.

In addition, the residual's behavior depends on the regularity of the maps $\btheta \mapsto \bG(\btheta)$ and $\btheta \mapsto \bG(\btheta)^+$. We assume both are Lipschitz continuous in the $\ell_2$ topology, with constants $C_{g}$ and $C_{g}^+$, respectively.

\vspace{-10pt}
\subsection{Radius of Convergence}\label{sec:main-result}

Define the $\ell_\eps$ ball $\B_{\infty}(\theta^\star, \eps) \coloneq \left\{\btheta : \norm{\btheta - \btheta^\star}_\infty \leq \eps \right\}$. As our main result, we establish the existence of a radius $\eps_0 > 0$, depending on the constants introduced in the previous section, such that $\B_{\infty}(\btheta^\star,\eps_0)$ is entirely contained within the strong basin of attraction of the solution for the loss of~\eqref{eq:projected-lstq}, provided the noise power $\norm{\bm{w}}_2^2$ is sufficiently small.

\begin{theorem}[Radius of Convergence for Projected Estimator] 
\label{thm:radius-projected}
    Define the constants
    \begin{align*}
            \alpha^\star &\coloneq \tfrac{\lambda_{\max,0}(C_\mu + 2pC_\Delta) + 2\Lambda_{\min, 1}(C_\mu + C_\Delta)}{\lambda_{\max,0}^2}, \\
            \beta^\star  &\coloneq \tfrac{\lambda_{\min,0}(C_\mu + C_\Delta) + \lambda_{\max,2}(C_\mu + 2C_\Delta)}{2\sqrt{\lambda_{\min,0}^{3}\lambda_{\min,2}}}, \\
            \gamma^\star &\coloneq \tfrac{C_{g}C_{g}^+(1 + \Lambda_{\min,0})\sqrt{\lambda_{\max,2}}}{\sqrt{N\lambda_{\min,0}\Lambda_{\min, 0}}},
        \end{align*}
    where, for $a \in \{0,1,2\}$,
    {\small
    \begin{align*}
        \lambda_{\min, a} &\coloneqq \tfrac{1}{2} \min_{i \in \llbracket p \rrbracket} \Big\{ \mu_{a}(\theta_i^\star, \theta_i^\star, \Delta) - \C_{a}(\theta_i^\star, \theta_i^\star, \Delta) \Big\}, \\
        \lambda_{\max, a} &\coloneqq \max_{i \in \llbracket p \rrbracket} \Big\{ \mu_{a}(\theta_i^\star, \theta_i^\star, \Delta) + \C_{a}(\theta_i^\star, \theta_i^\star, \Delta) \Big\}, \\
        \Lambda_{\min, a} &\coloneqq \tfrac{1}{2} \min_{i \in \llbracket p \rrbracket} \mu_{a}(\theta_i^\star, \theta_i^\star, \Delta) - \mathcal{S}_a(\btheta^\star), \\
        \Lambda_{\max, a} &\coloneqq \max_{i \in \llbracket p \rrbracket} \mu_{a}(\theta_i^\star, \theta_i^\star, \Delta) + \mathcal{S}_a(\btheta^\star).
    \end{align*}
    }
    with $\mathcal{S}_a(\btheta) \coloneq \max_{i \in \llbracket p \rrbracket}\sum_{j = 1}^p\C_{a}(\theta_i, \theta_j, \Delta)$.
    Assume 
    \begin{equation}\label{eq:feasability}
        \norm{\bm{w}}_2 < \tfrac{\Lambda_{\min, 1}}{\lambda_{\max,0}}\sqrt{\tfrac{\lambda_{\min,0}}{\lambda_{\max,2}}} \norm{\bx}_2,
    \end{equation}
    and set
    \begin{equation}\label{eq:radius-projected}
        \eps_0 \coloneq \frac{\tfrac{\Lambda_{\min, 1}}{\lambda_{\max,0}}\norm{\bx}_2^2 - \sqrt{\tfrac{\lambda_{\max,2}}{\lambda_{\min,0}}}\norm{\bx}_2\norm{\bm{w}}_2}{\alpha^\star\norm{\bx}_2^2 + \beta^\star\norm{\bx}_2\norm{\bm{w}}_2 + \gamma^\star\norm{\bx^\star}_2} > 0.
    \end{equation}
    Then, for any $\eps \in [0, \eps_0)$, there exists $N_0 \in \N$ such that if $N \geq N_0$ then $\B_{\infty}(\btheta^\star, \eps)$ is a strong basin of attraction for the loss of~\eqref{eq:projected-lstq}.
\end{theorem}
Theorem~\ref{thm:radius-projected} shows that the radius $\eps_0$ is controlled by the block minimum eigenvalue $\Lambda_{\min,1}$ and the dynamic range $\kappa_0 = \lambda_{\max,0}/\lambda_{\min,0}$. Large values of $\Lambda_{\min,1}$ arise from PSFs whose first derivatives have large coherence $\mu_{1}(\theta_i^\star, \theta_i^\star, 0) $, while small a dynamic range $\kappa_0$ can be achieved with sufficient separation $\Delta$.

Equation~\eqref{eq:radius-projected} further indicated that the radius shrinks as the noise power increases. The lower bound vanishes when the condition~\eqref{eq:feasability} is violated, hence no theoretical guarantees can be given in that case. Finally, the higher-order constants $\alpha^\star,\beta^\star,\gamma^\star$ quantify the Lipschitz sensitivity of the blocks, with $\gamma^\star$ vanishing as $N$ grows, meaning larger sample sizes are beneficial.

\section{Proof Elements of Theorem 1}\label{sec:proof}

We outline the key proof steps. Complete derivations are omitted due to space limitations.  
The Hessian of~\eqref{eq:projected-lstq} reads~\cite{projected-hessian}
\begin{align}\label{eq:hessian}
    \nabla^2\L(\btheta) &= \tfrac{1}{N}\Big( \diag{\bG_0^+\bx}^\top\bG_1^\top\bG_1\diag{\bG_0^+\bx} \nonumber \\
    &\quad +  (\Id_p \otimes \P_{\bG(\btheta)}^\perp\bx)^\top\diag{\bG_2}\diag{\bG_0^+\bx} \Big) \nonumber \\
    &\eqqcolon \tfrac{1}{N}\left(\E + \bR\right),
\end{align}
where $\otimes$ denotes the Kronecker product. For $\bm{B} \in \R^{N \times M}$, $\diag{\bm{B}}$ maps $\bm{B}$ into a matrix of size $pN \times M$ supported by $p$ diagonal blocks of size $N \times M_i$. Observe~\eqref{eq:hessian} splits the Hessian matrix into a curvature term $\E$ and a residual  $\bR$. Weyl's inequality yields
\begin{subequations}
    \begin{align}
        \lmin(\nabla^2\L(\btheta)) &\geq \tfrac{1}{N}\big(\lmin(\E) - \lmax(\bR)\big), \label{eq:min-E}
    \end{align}
\end{subequations}
In the rest of the proof, we bound $\lmin(\E)$ and $\lmax(\bR)$ separately.

\begin{lemma}\label{thm:curvature}
    The matrix $\E$ satisfies
    \begin{subequations}
        \begin{align}
            \lmin(\E) &\geq \tfrac{\lmin(\bG_1^\top\bG_1)}{\max_i \lmax(\bG_{0,i}^\top\bG_{0,i})}\norm{\bx}_2^2.
        \end{align}
\end{subequations}
\end{lemma}

To control $\lmax(\bm{R})$, we harness its diagonal structure of the matrix $\bR$ with
\begin{multline}\label{eq:residual-projected} 
\lmax(\bm{R}) = \max_{i \in \llbracket p \rrbracket}|\langle \bG_{2,i}\bG_{0,i}^+\bx, \P_{\bG}^\perp \bx\rangle|  \\ \leq \max_{i \in \llbracket p \rrbracket}  \sqrt{\tfrac{\lmax(\bG_{2,i}^\top\bG_{2,i})}{\lmin(\bG_{0,i}^\top\bG_{0,i})}}\norm{\bx}_2\left( \norm{\P_{\bG}^\perp\bx^\star}_2 + \norm{\bm{w}}_2 \right),
\end{multline}  
where $\bG_a \coloneqq [\bG_{a,1}, \dots, \bG_{a,p}] \in \R^{N \times M}$ with 
$\bG_{a,i} \coloneqq [\partial_a \bg(\theta_i, t_{i,1}), \dots, \partial_a \bg(\theta_i, t_{i,M_i})] \in \R^{N \times M_i}$.

\begin{lemma}\label{thm:eps-residual}
    Let $\btheta \in \B_{\infty}(\btheta^\star,\eps)$. Then
    \begin{align}\label{eq:eps-residual-projected}
        \MoveEqLeft[0] \norm{\P_{\bG}^\perp \bx^\star}_2 \leq \norm{\bx^\star}_2\Bigg[ NC_{g}^+C_{g}^2 \eps^2 & \nonumber \\ 
        &\hspace{12pt} + \left. \sqrt{N}\left( \tfrac{C_{g}}{\sqrt{\lmin(\bG_0^{\star^\top}\bG_0^\star)}} + \norm{\bG_0^\star}_2C_{g}^+C_{g} \right)\eps  \right].
    \end{align}
\end{lemma}

Substituting inequality~\eqref{eq:eps-residual-projected}  into~\eqref{eq:residual-projected} provides a bound on the residual matrix $\lmax(\bm{R})$ in terms of the spectrum of Gramian matrices of the form $\bm{G}_a^\top \bm{G}_a$. 
To explicate those Gramians in \eqref{eq:residual-projected}, we further exploit coherence and total coherence using the following lemma. 
\begin{lemma}[Controlling the spectra of Gramians]\label{thm:gramians}
    For any $\btheta \in \Theta^p$ and all $a \in \{0,1,2\}$, the following bounds hold:
    \begin{subequations}\label{eq:coherence_bounds}
        \begin{align}
            \lmin(\bG_{a,i}^\top\bG_{a,i}) &\geq \tfrac{1}{2} \mu_{a}(\theta_i, \theta_i, 0) - \C_{a}(\theta_i, \theta_i, \Delta), \label{eq:single-block-min-eigval} \\
            \lmax(\bG_{a,i}^\top\bG_{a,i}) &\leq \mu_{a}(\theta_i, \theta_i, 0) + \C_{a}(\theta_i, \theta_i, \Delta), \label{eq:single-block-max-eigval} \\
            \lmin(\bG_{a}^\top\bG_{a}) &\geq \tfrac{1}{2} \min_{i \in \llbracket p \rrbracket}\mu_{a}(\theta_i, \theta_i, 0) - \mathcal{S}_a(\btheta) \label{eq:block-min-eigval},\\
            \lmax(\bG_{a}^\top\bG_{a}) &\leq \max_{i \in \llbracket p \rrbracket}\mu_{a}(\theta_i, \theta_i, 0) + \mathcal{S}_a(\btheta). \label{eq:block-max-eigval}
        \end{align}
    \end{subequations}
\end{lemma}
Finally, we exploit the Lipschitz continuity assumption on the coherence and total coherence to derive bound on the quantities~\eqref{eq:coherence_bounds} that depend only on the coherence and total coherence of ground truth parameters $\bm{\theta}^\star$. This
yields a smooth envelopes $\eps \mapsto \hat{\xi}(\eps)$ satisfying
\(    \xi(\eps) \leq \lmin(\nabla^2\L(\btheta)),
\)
which in turn yields the explicit lower bound strong basin radius stated in Theorem~\ref{thm:radius-projected}.

\vspace{-10pt}
\section{Numerical Experiments}\label{sec:experiments}
\vspace{-5pt}
In this section, we present numerical experiments designed to illustrate how the choice of parametric family influences the behavior of the problem. We focus on the $u$-Laplace kernel, defined for $u > 0$ by $g(\theta, t) = \exp\left( -\left(|t|/\theta\right)^u \right)$, where the parameter $u$ controls the decay rate of the kernel tails.

\begin{figure}[t]
    \centering
    \begin{tikzpicture}
        \node (image) at (0,0) {\includegraphics[width=0.95\linewidth]{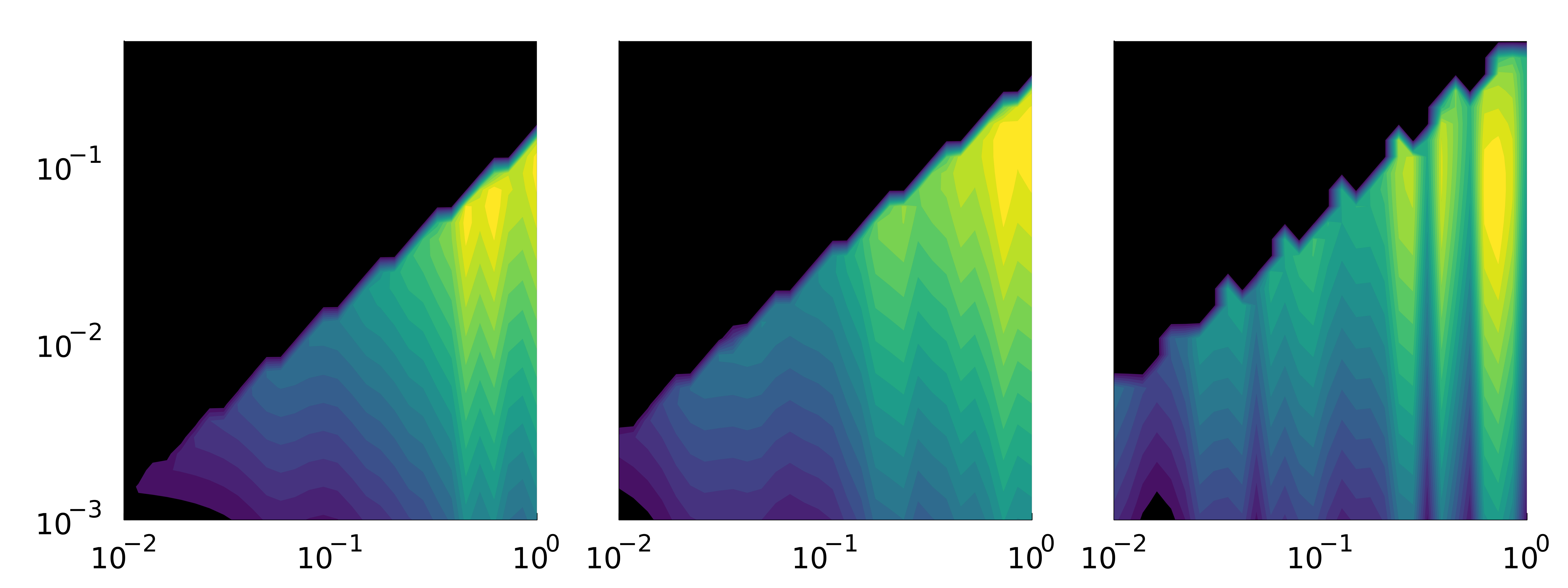}};
        \node[above=of image, yshift=-1.3cm, xshift=-2.6cm , scale=0.7] {$\begin{array}{c}
             u = 1 \\
             \eps_{\max} = 1.36\cdot10^{-3}
        \end{array}$};
        \node[above=of image, xshift=0.2cm, yshift=-1.3cm, scale=0.7] {$\begin{array}{c}
             u = 2 \\
             \eps_{\max} = 3.53\cdot10^{-3}
        \end{array}$};
        \node[above=of image, xshift=3cm, yshift=-1.3cm, scale=0.7] {$\begin{array}{c}
             u = 20 \\
             \eps_{\max} = 5.09\cdot10^{-3}
        \end{array}$};

        \node[left=of image, scale=0.7, xshift=1.8cm] {$\theta^\star$};
        \node[below=of image, xshift=0.2cm, yshift=1.2cm, scale=0.7] {$\Delta$};
        \node[below=of image, xshift=3cm, yshift=1.2cm, scale=0.7] {$\Delta$};
        \node[below=of image, xshift=-2.6cm, yshift=1.2cm, scale=0.7] {$\Delta$};
    \end{tikzpicture}
    \vspace{-10pt}
    \caption{Normalized $\eps_0$ from~\eqref{eq:radius-projected} for $u$-Laplace kernels with $u=1$ (left), $u=2$ (center), and $u=20$ (right). Black regions correspond to non-positive radii (ill-posed cases).}
    \label{fig:radii}
\end{figure}
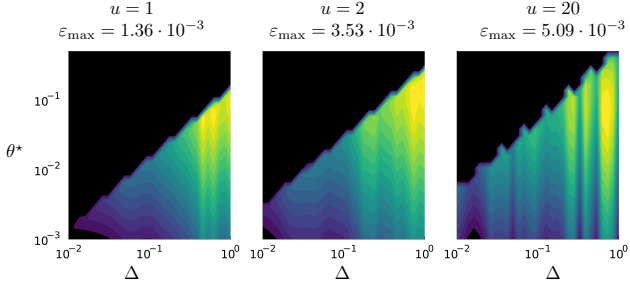

\vspace{-10pt}
\paragraph*{Evaluation of the lower bound~\eqref{eq:radius-projected}.} We first evaluate the lower bound on the radius of the strong convexity region~\eqref{eq:radius-projected} predicted by Theorem~\ref{thm:radius-projected} in the noiseless regime. Figure~\ref{fig:radii} presents the radii computed over $(\theta^\star,\Delta) \in [10^{-3},1]\times[10^{-2},1]$ for three different PSF classes with $u\in\{1, 2, 20\}$ and normalized by the global maximum. Black regions mark ill-posed settings where the condition of Theorem~\ref{thm:radius-projected} is not met. It shows that a larger $u$ enlarges the well-posed region and yields bigger radii, indicating that fast-decaying kernels fundamentally provide more stable optimization geometries.   

\vspace{-10pt}
\paragraph*{Monte-Carlo validation of Theorem~\ref{thm:radius-projected}.}\label{sec:monte-carlo}

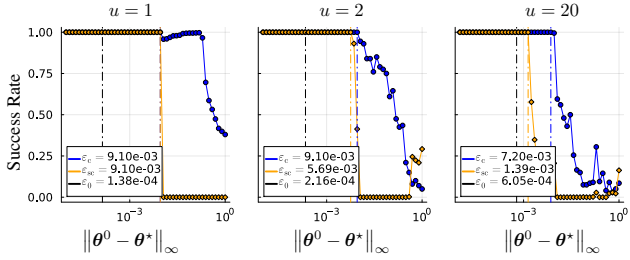
\begin{figure}[t]
    \centering
    \begin{tikzpicture}
        \node (image) at (0,0) {\includegraphics[width=0.95\linewidth]{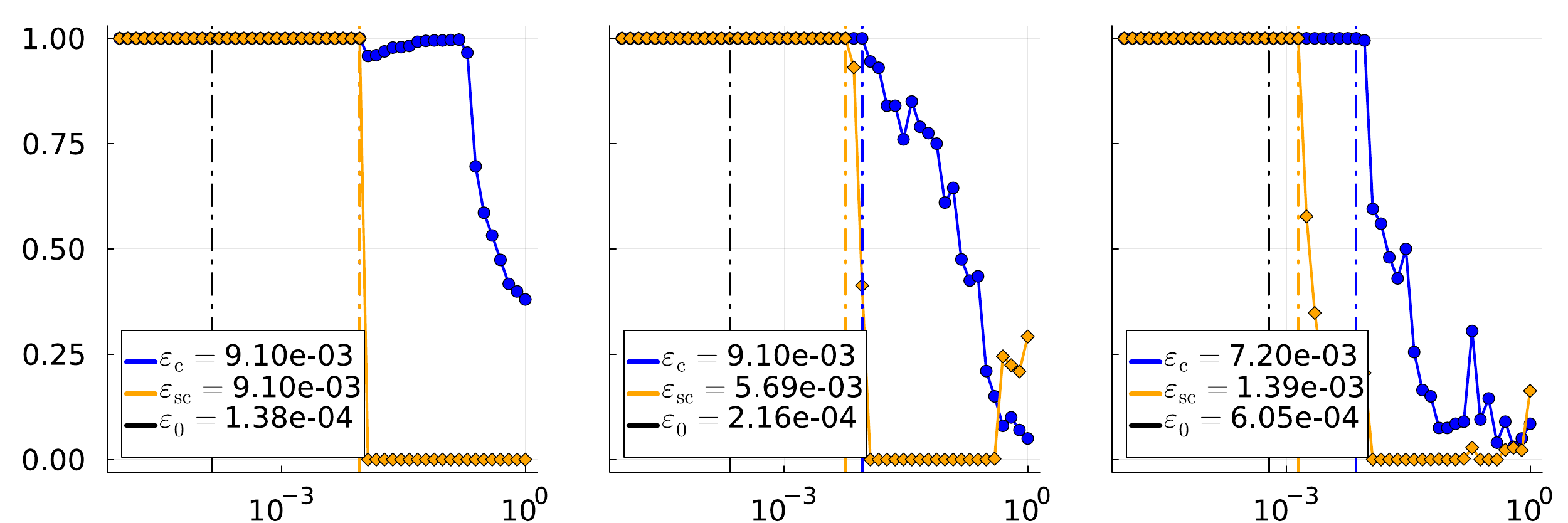}};
        \node[above=of image, yshift=-1.2cm, xshift=-2.6cm , scale=0.7] {$u = 1$};
        \node[above=of image, xshift=0.2cm, yshift=-1.2cm, scale=0.7] {$u=2$};
        
        \node[above=of image, xshift=3cm, yshift=-1.2cm, scale=0.7] {$u = 20$};

        \node[left=of image, scale=0.7, xshift=1.5cm, yshift=1cm, rotate=90] {Success Rate};

        \node[below=of image, xshift=0.2cm, yshift=1.1cm, scale=0.7] {$\norm{\btheta^0 - \btheta^\star}_\infty$};
        \node[below=of image, xshift=3cm, yshift=1.1cm, scale=0.7] {$\norm{\btheta^0 - \btheta^\star}_\infty$};
        \node[below=of image, xshift=-2.6cm, yshift=1.1cm, scale=0.7] {$\norm{\btheta^0 - \btheta^\star}_\infty$};
    \end{tikzpicture}
    \vspace{-20pt}
    \caption{Monte-Carlo success rates for the strong-convexity and convergence tests across initial distances $\|\btheta^0-\btheta^\star\|_\infty$.}
    \label{fig:monte_carlo}
\end{figure}

We compare the theoretical radius with empirical behavior by probing strong convexity and convergence as a function of the initialization distance $\eps=\|\btheta^0-\btheta^\star\|_\infty$. The signal $x(\cdot)$ consists of two groups of three spikes over $I=[-1,1]$, sampled with $N=10^4$ at SNR $10$~dB. We set $\btheta^\star=[10^{-2},10^{-2}]$, $\Delta=0.4$, amplitudes $\eta^\star_{i,\ell}=1/\|\sum_{k,\ell}\bg_{k,\ell}\|_2$, and use $u$-Laplace kernels with $u\in\{1,2,20\}$. Figure~\ref{fig:monte_carlo} reports the proportion of $100$ trials where the loss is strongly convex (orange) and where the method converges (blue). The vertical markers indicate the estimated convergence radius $\eps_\mathrm{c}$, the strong convexity radius $\eps_\mathrm{sc}$, and the theoretical bound $\eps_0$ from Theorem~\ref{thm:radius-projected}. For $u=1$ and $u=2$, the bound $\eps_0$ is conservative, about one order of magnitude below the empirical radii. For $u=20$, the bound becomes sharp, showing that fast-decaying kernels not only enlarge the basin of attraction but also make the theoretical radius predictive.

\begin{figure}[t]
    \centering
    \begin{tikzpicture}
        \node (mse) at (0,0) {\includegraphics[width=0.95\linewidth]{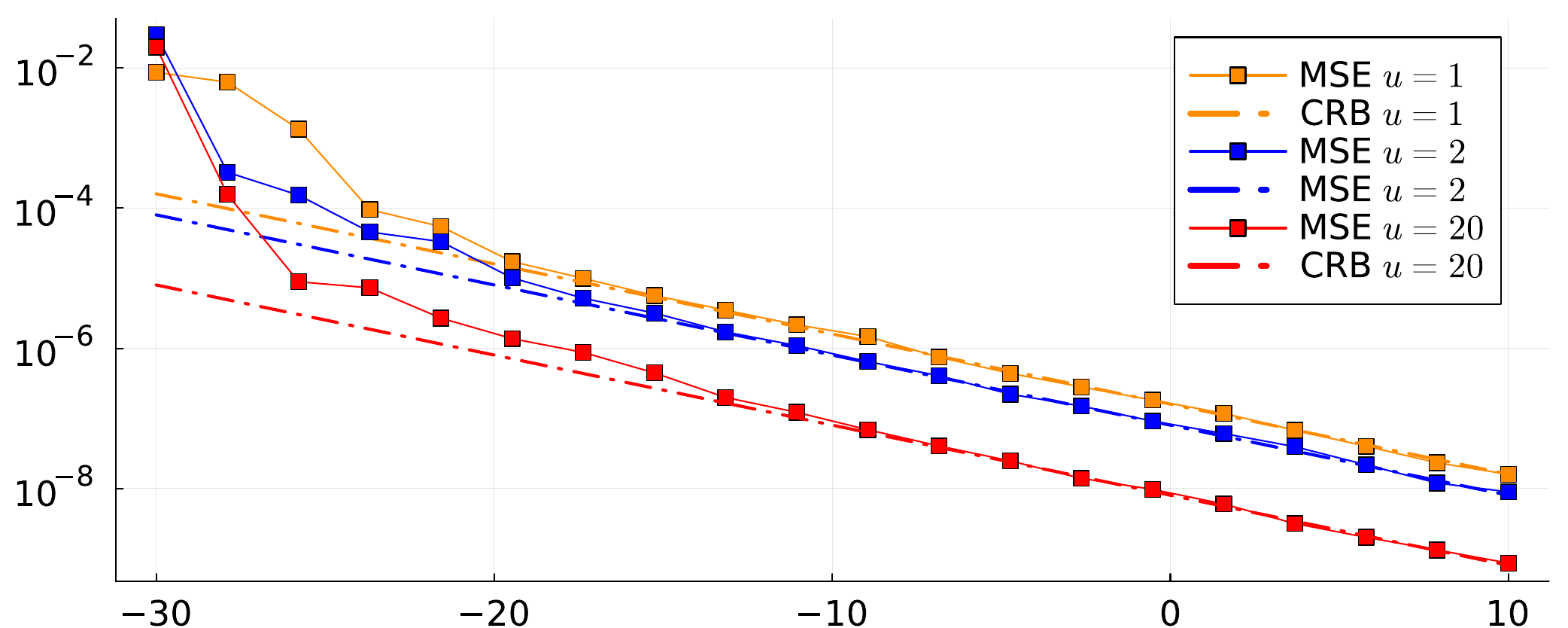}};

        \node[below=of mse,  yshift=1.1cm, scale=0.7] {SNR [dB]};

        \node[left=of mse, xshift=1cm, yshift=1.1cm, scale=0.7, rotate=90]{Mean squared error};     
    \end{tikzpicture}
    \vspace{-20pt}
    \caption{Mean squared error (solid) and Cramér–Rao bound (dashed) vs. SNR for different $u$-Laplace kernels.}
    \label{fig:mse}
\end{figure}

\vspace{-5pt}
\paragraph*{Performance vs SNR.}\label{sec:snr}
Assuming $\bm{w}$ is white Gaussian noise, we estimate the MSE of the recovered $\hat{\btheta}$ over 100 trials alongside the Cramér–Rao bound (CRB). Figure~\ref{fig:mse} shows that both the MSE and the CRB shrink significantly as $u$ increases. Hence, fast-decaying kernels improve not only theoretical radii and basin estimates, but also recovery accuracy under noise.

\vspace{-5pt}
\paragraph*{Application to LIBS.}\label{sec:libs}

Finally, we evaluate the method on real-world LIBS data. A spectrum consists of emission lines at known wavelengths, broadened by plasma interactions, the experimental setup, and sample composition. The goal is to recover elemental concentrations from the observed spectrum. In our model formulation, $\theta_i$ denotes the line-shape parameter of species $i$, and $\eta_{i,\ell}$ the integrated intensity of its $\ell$-th emission line. The recovered intensities are then fed into the Boltzmann equations~\cite{griem1974spectral} to yield a concentration estimate. Unlike existing approaches, our method jointly recovers line widths and intensities, removing the need for \emph{a priori} width estimation required in calibration-free LIBS~\cite{cflibs, review_cf_libs}. We test on a calibrated Aluminum alloy composed of $91.50\%$ Al, $5.52\%$ Cu, $0.98\%$ Fe, and $0.39\%$ Mg. An overall relative fit error of $6.83\%$ is obtained. Concentrations are recovered with relative errors of $5.42\%$, $46.67\%$, $9.41\%$, and $32.52\%$, respectively. The relatively higher Cu and Mg errors reflect their weaker emissivity, while Fe is accurately estimated despite low abundance owing to its strong emission.

\begin{figure}[t]
    \centering
    \begin{tikzpicture}
        \node (mse) at (0,0) {\includegraphics[width=0.95\linewidth]{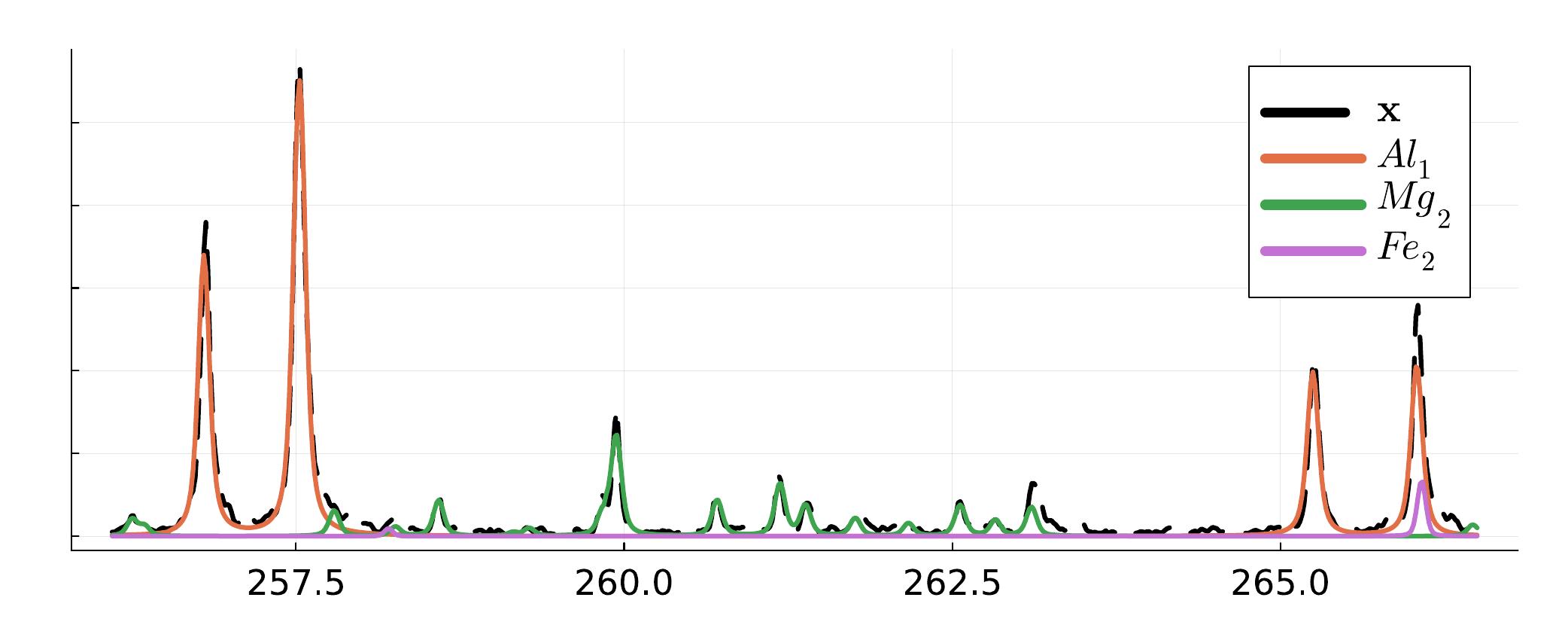}};

        \node[below=of mse,  yshift=1.1cm, scale=0.7] {$\lambda$ [nm]};

        \node[left=of mse, xshift=1cm, yshift=1.1cm, scale=0.7, rotate=90]{Line Intensity [a. u.]};     
    \end{tikzpicture}
    \vspace{-20pt}
    \caption{Details of recovered spectrum in the $[256.1, 266.5]$~nm range. Recovered individual species in a range of colors.}
    \label{fig:spectrum}
\end{figure}

\section{Conclusion}\label{sec:conclusion}

In this paper, we leveraged variable projection to formulate an improved estimator for the PSF unmixing problem. In Theorem~\ref{thm:radius-projected}, we provided a lower bound on the size of the strong basin of attraction in terms of problem parameters. We conducted numerous numerical experiments that underpin the governing factor in problem difficulty corresponds to kernel tail decay speed, which corroborates our theoretical findings. We demonstrated real-world applicability of our method in LIBS spectroscopy, where it recovers elemental concentrations without prior width calibration. Future work includes establishing theoretical guarantees for algorithms yielding solutions sparse in~$\mathcal{T}$, extending our recent results~\cite{michelena:gretsi2025}.

\clearpage

\bibliographystyle{IEEEbib}
\bibliography{icassp_refs}

\begin{thebibliography}{10}

\bibitem{huang2008three}
Bo~Huang, Wenqin Wang, Mark Bates, and Xiaowei Zhuang,
\newblock ``Three-dimensional super-resolution imaging by stochastic optical reconstruction microscopy,''
\newblock {\em Science}, vol. 319, no. 5864, pp. 810--813, 2008.

\bibitem{huang2018single}
Yuanfei Huang, Jie Li, Xinbo Gao, Lihuo He, and Wen Lu,
\newblock ``Single image super-resolution via multiple mixture prior models,''
\newblock {\em IEEE Transactions on Image Processing}, vol. 27, no. 12, pp. 5904--5917, 2018.

\bibitem{knudson2014inferring}
Karin~C Knudson, Jacob Yates, Alexander Huk, and Jonathan~W Pillow,
\newblock ``Inferring sparse representations of continuous signals with continuous orthogonal matching pursuit,''
\newblock {\em Advances in neural information processing systems}, vol. 27, 2014.

\bibitem{li2014sparse}
Yuanqing Li, Zhu~Liang Yu, Ning Bi, Yong Xu, Zhenghui Gu, and Shun-ichi Amari,
\newblock ``Sparse representation for brain signal processing: a tutorial on methods and applications,''
\newblock {\em IEEE Signal Processing Magazine}, vol. 31, no. 3, pp. 96--106, 2014.

\bibitem{applebaum2012asynchronous}
Lorne Applebaum, Waheed~U Bajwa, Marco~F Duarte, and Robert Calderbank,
\newblock ``Asynchronous code-division random access using convex optimization,''
\newblock {\em Physical Communication}, vol. 5, no. 2, pp. 129--147, 2012.

\bibitem{chi2013compressive}
Yao Xie, Yuejie Chi, Lorne Applebaum, and Robert Calderbank,
\newblock ``Compressive demodulation of mutually interfering signals,''
\newblock in {\em 2012 IEEE Statistical Signal Processing Workshop (SSP)}, 2012, pp. 592--595.

\bibitem{review_cf_libs}
Zhenlin Hu, Deng Zhang, Weiliang Wang, Feng Chen, Yubin Xu, Junfei Nie, Yanwu Chu, and Lianbo Guo,
\newblock ``A review of calibration-free laser-induced breakdown spectroscopy,''
\newblock {\em TrAC Trends in Analytical Chemistry}, vol. 152, pp. 116618, 2022.

\bibitem{li2016identifiability}
Yanjun Li, Kiryung Lee, and Yoram Bresler,
\newblock ``Identifiability in blind deconvolution with subspace or sparsity constraints,''
\newblock {\em IEEE Transactions on information Theory}, vol. 62, no. 7, pp. 4266--4275, 2016.

\bibitem{romberg2010sparse}
Justin Romberg and Ramesh Neelamani,
\newblock ``Sparse channel separation using random probes,''
\newblock {\em Inverse Problems}, vol. 26, no. 11, pp. 115015, 2010.

\bibitem{chi_guaranteed_2016}
Yuejie Chi,
\newblock ``Guaranteed blind sparse spikes deconvolution via lifting and convex optimization,''
\newblock vol. 10, no. 4, pp. 782--794.

\bibitem{Chretien2020}
S{\'e}bastien Chr{\'e}tien and Himanshu Tyagi,
\newblock ``Multi-kernel unmixing and super-resolution using the modified matrix pencil method,''
\newblock {\em Journal of Fourier Analysis and Applications}, vol. 26, no. 1, pp. 18, 2020.

\bibitem{traonmilin2024strong}
Yann Traonmilin, Jean-Fran{\c{c}}ois Aujol, Pierre-Jean B{\'e}nard, and Arthur Leclaire,
\newblock ``On strong basins of attractions for non-convex sparse spike estimation: upper and lower bounds,''
\newblock {\em Journal of Mathematical Imaging and Vision}, vol. 66, no. 1, pp. 57--74, 2024.

\bibitem{costa_local_2023}
Maxime Ferreira Da~Costa and Yuejie Chi,
\newblock ``Local geometry of nonconvex spike deconvolution from low-pass measurements,''
\newblock {\em IEEE Journal on Selected Areas in Information Theory}, vol. 4, pp. 1--15, 2023.

\bibitem{gabet2025global}
Joseph Gabet, Meghna Kalra, Maxime Ferreira Da~Costa, and Kiryung Lee,
\newblock ``Global convergence of {ESPRIT} with preconditioned first-order methods for spike deconvolution,''
\newblock {\em arXiv preprint arXiv:2502.08035}, 2025.

\bibitem{mars-spectrum-fitting}
P.B. Hansen, S.~Schröder, S.~Kubitza, K.~Rammelkamp, D.S. Vogt, and H.-W. Hübers,
\newblock ``Modeling of time-resolved libs spectra obtained in martian atmospheric conditions with a stationary plasma approach,''
\newblock {\em Spectrochimica Acta Part B: Atomic Spectroscopy}, vol. 178, pp. 106115, 2021.

\bibitem{michelena}
Santos Michelena, Maxime~Ferreira Da~Costa, and José Picheral,
\newblock ``Convergence guarantees for unmixing {PSFs} over a manifold with non-convex optimization,''
\newblock in {\em 2025 IEEE Statistical Signal Processing Workshop (SSP)}, 2025, pp. 161--165.

\bibitem{variableProjection}
G.~H. Golub and V.~Pereyra,
\newblock ``The differentiation of pseudo-inverses and nonlinear least squares problems whose variables separate,''
\newblock {\em SIAM Journal on Numerical Analysis}, vol. 10, no. 2, pp. 413--432, 1973.

\bibitem{varpro-meta-review}
V.~Pereyra G.~H.~Golub,
\newblock ``Separable nonlinear least squares: the variable projection method and its applications,''
\newblock {\em Inverse Problems}, vol. 19, no. 2, pp. R1, feb 2003.

\bibitem{projected-hessian}
Axel Ruhe and Per~{\AA}ke Wedin,
\newblock ``Algorithms for separable nonlinear least squares problems,''
\newblock {\em SIAM Review}, vol. 22, no. 3, pp. 318--337, 1980,
\newblock Accessed: 2025-06-25.

\bibitem{griem1974spectral}
Hans~R. Griem,
\newblock {\em Spectral Line Broadening by Plasmas},
\newblock Academic Press, New York, 1974.

\bibitem{cflibs}
A.~Ciucci, M.~Corsi, V.~Palleschi, S.~Rastelli, A.~Salvetti, and E.~Tognoni,
\newblock ``New procedure for quantitative elemental analysis by laser-induced plasma spectroscopy,''
\newblock {\em Applied Spectroscopy}, vol. 53, no. 8, pp. 960--964, 1999.

\bibitem{michelena:gretsi2025}
Santos Michelena, Jos{\'e} Picheral, and Maxime Ferreira Da~Costa,
\newblock ``Reconstruction de r{\'e}ponses impulsionnelles sur une vari{\'e}t{\'e} pour la spectroscopie de plasma induit par laser,''
\newblock in {\em Proc. GRETSI 2025}, Jul 2025.

\end{thebibliography}

\end{document}